\documentclass[runningheads]{llncs}
\usepackage{tikz}
\usetikzlibrary{positioning}
\usepackage[T1]{fontenc}
\usepackage{graphicx}
\usepackage{amsmath,amssymb}
\usepackage{booktabs}
\usepackage{multirow}
\usepackage{url}
\usepackage{xcolor}
\usepackage{enumitem}
\usepackage{microtype}
\usepackage{subcaption}

\newcommand{\attr}[1]{\texttt{#1}}

\begin{document}
\title{Real Talk, Virtual Faces: Symbolic--Semantic Discourse Geometry of Virtual and Human Influencer Audiences}
\titlerunning{Symbolic--Semantic Discourse Geometry}
%
%
\author{Shahram Chaudhry \and Sidahmed Benabderrahmane \and Talal Rahwan}
\authorrunning{S. Chaudhry et al.}
%
\institute{New York University (NYUAD), Division of Science, Computer Science Department\\
\email{sc9425@nyu.edu}}
\maketitle              

\begin{abstract}
Virtual influencers~(VIs)---digitally constructed social-media personas---are
becoming prominent actors in online culture, marketing, and identity formation.
Yet it remains unclear whether audiences respond to them through the same
behavioural discourse patterns used for human influencers~(HIs), or whether
virtuality gives rise to distinctive modes of reaction. Existing studies often
rely on surveys, engagement statistics, or marginal sentiment distributions,
which reveal what audiences say but not how affective, topical, and
psycholinguistic signals are jointly organised.

We introduce a symbolic--semantic framework for analysing audience discourse
around virtual and human influencers. First, we use Formal Concept Analysis and
association rule mining to extract closed co-occurrence structures from
sentiment labels, topic tags, and Big Five psycholinguistic cues. Second, we
render the extracted formal concepts as natural-language descriptions, embed
them using MiniLM, and compare their semantic geometry across virtual and human
influencer audiences.

Applied to 69,498 YouTube comments from three matched VI--HI influencer pairs,
our analysis reveals three findings. First, HI discourse is organised around a
compact, stability-centred pattern in which low neuroticism anchors positive
sentiment, whereas VI discourse supports multiple discourse regimes. Second,
VI concepts are more semantically dispersed than HI concepts, while both groups
show strong symbolic--semantic alignment between closed-set structure and
embedding geometry. Third, VI discourse contains a distinct artificial-identity
region and a higher concentration of negative sentiment in psychologically
sensitive topics such as mental health, body image, and artificial identity.
Together, these findings suggest that virtuality reshapes not only the sentiment
of audience reactions, but also the symbolic and semantic organisation of online
social discourse.
\end{abstract}

\section{Introduction}

Virtual influencers~(VIs)---synthetic personas designed to look, speak, and
interact like social-media creators---are increasingly visible in online culture,
advertising, music, fashion, and youth-oriented digital communities. Their rise
raises a computational social science question: do audiences respond to virtual
figures through the same discourse patterns used for human influencers~(HIs), or
does virtuality reorganise how users combine emotion, topical attention, and
social judgement?

Prior work on virtual influencers has largely relied on surveys, experiments,
or aggregate engagement statistics. These approaches are valuable, but they
usually analyse individual signals in isolation: sentiment scores, authenticity
perceptions, trust, or engagement counts. Such marginal summaries can miss the
structure of audience behaviour. For example, two influencer groups may show
similar prevalence of appearance-related comments, yet differ sharply in how
appearance co-occurs with sentiment, psycholinguistic style, and identity
scrutiny. This motivates a shift from asking only \emph{what} audiences say to
asking \emph{how} multiple behavioural signals are organised.

We address this problem through a symbolic--semantic discourse analysis
framework. The symbolic layer uses Formal Concept Analysis~(FCA) and association
rule mining to identify closed co-occurrence patterns among sentiment labels,
topic tags, and Big Five psycholinguistic cues. FCA is useful here because its
formal concepts capture maximal bundles of attributes shared by groups of
comments or weeks, revealing co-occurrence structure that cannot be read from
marginal frequencies alone~\cite{GanterWille1999}. The semantic layer then
renders these formal concepts as short natural-language descriptions, embeds
them using MiniLM, and analyses the resulting concept geometry. This allows us
to test whether symbolic closed-set structure corresponds to coherent semantic
regions and whether VI discourse occupies a broader or more differentiated
semantic space than HI discourse.

A motivating example is \attr{topic\_appearance}. In our data, appearance has
near-equal marginal prevalence in VI and HI comments, yet it participates in
multiple VI rule clusters and in none of the HI rule sets. This is precisely the
kind of behavioural asymmetry that marginal statistics obscure but
symbolic--semantic structure can reveal.

\paragraph{Contributions.}
We make four contributions:
\begin{enumerate}[leftmargin=*]
\item We construct a pair-matched YouTube comment dataset of 69,498 comments
      from three virtual--human influencer pairs, enriched with sentiment
      labels, topic tags, and psycholinguistic style cues.
\item We introduce a symbolic--semantic discourse analysis pipeline that
      combines closed-set mining, association rules, MiniLM concept embeddings,
      and cluster-level semantic interpretation.
\item We show that VI and HI audiences differ not only in marginal sentiment
      and topic prevalence, but in the organisation of their discourse: VI
      comments yield a denser and more diverse rule grammar, including an
      appearance-discourse pattern absent from HI despite similar appearance
      prevalence.
\item We compare symbolic and semantic concept geometries, showing strong
      symbolic--semantic alignment for both groups, but higher semantic
      dispersion for VI discourse and a VI-specific artificial-identity cluster.
\end{enumerate}

\paragraph{Research questions.}
\begin{enumerate}[leftmargin=*,label=\textbf{RQ\arabic*.}]
\item \textbf{(Audience discourse structure.)} How do the co-occurrence
      patterns of sentiment, topic, and psycholinguistic cues differ between
      VI and HI audiences?
\item \textbf{(Symbolic--semantic geometry.)} Do symbolic discourse structures
      extracted from formal concepts correspond to coherent semantic regions in
      embedding space, and how does this geometry differ between VI and HI
      audiences?
\item \textbf{(Virtuality-sensitive topics.)} How do audience reactions around
      artificial identity, authenticity, body image, and mental health differ
      between virtual and human influencer contexts?
\end{enumerate}

\paragraph{Paper organisation.}
Section~2 presents the conceptual framework. Section~3 reviews related work on
virtual influencers, social-media discourse analysis, and symbolic--semantic
interpretability. Section~4 describes the data and annotation pipeline.
Section~5 presents the symbolic--semantic methodology. Section~6 reports the
empirical results, and Section~7 discusses their implications for computational
social science and AI-mediated social interaction.

\section{Conceptual Framework}

We conceptualise virtual influencers as \emph{synthetic social actors}: mediated
personas that perform the interactional role of influencers while foregrounding
questions of artificiality, embodiment, authenticity, and social presence. From a
computational social science perspective, the key question is not only whether
comments about VIs are more positive or negative than comments about HIs, but
whether audiences organise their reactions through different combinations of
affect, topic, and psycholinguistic style.

Figure~\ref{fig:conceptual_framework} summarises the analytical logic. The
starting point is the socio-technical condition of \emph{virtuality}: the audience
knows, infers, or debates that the visible persona is partly or fully synthetic.
This condition may activate interpretive frames around authenticity, artificial
identity, appearance, and emotional credibility. These frames become observable
in comment discourse through combinations of sentiment, topic, and
psycholinguistic style cues. Our goal is therefore to model \emph{discourse
organisation}: the way these signals co-occur, form closed symbolic structures,
and occupy semantic regions.

\begin{figure}[!t]
\centering
\fbox{\begin{minipage}{0.92\linewidth}
\centering
\vspace{1mm}
\textbf{Virtuality as socio-technical condition}\\[1mm]
$\Downarrow$\\[-1mm]
\textit{Artificial identity, authenticity ambiguity, synthetic embodiment}\\[1mm]
$\Downarrow$\\[-1mm]
\textbf{Audience discourse organisation}\\[1mm]
\textit{sentiment $\times$ topic $\times$ psycholinguistic style}\\[1mm]
$\Downarrow$\\[-1mm]
\textbf{Symbolic--semantic discourse regimes}\\[1mm]
\textit{closed concepts, association rules, semantic clusters}
\vspace{1mm}
\end{minipage}}
\caption{Conceptual framework. We treat virtuality not as a directly identifiable
causal treatment, but as a socio-technical condition under which audiences may
interpret identity, authenticity, embodiment, and affective credibility
differently. The empirical analysis therefore focuses on discourse organisation:
how affective, topical, and psycholinguistic signals co-occur in VI and HI
audience responses.}
\label{fig:conceptual_framework}
\end{figure}

This framing leads to two methodological commitments. First, we avoid reducing
audience response to marginal sentiment or topic prevalence alone, because such
summaries cannot show how signals combine. Second, we avoid interpreting
associations as causal effects of virtuality. The matched VI--HI design supports
comparative structural analysis, but unobserved differences in creators, content,
audience composition, and platform dynamics may remain. We therefore interpret
our findings as pair-matched contrasts in discourse organisation.


\section{Related Work}

\subsection{Virtual influencers as synthetic social actors}

Research on virtual influencers consistently highlights authenticity, parasocial
relations, realism, and artificial identity as central to audience response.
Batista and Chimenti~\cite{Batista2021} show that highly humanised VIs can
produce emotionally complex reactions ranging from admiration to skepticism.
Arsenyan and Mirowska~\cite{Arsenyan2022} find that disclosure of virtual status
shapes trust and purchase intent, while Xie-Carson et al.~\cite{XieCarson2024}
connect VI reception to uncanny-valley dynamics in tourism contexts. Sands et
al.~\cite{Sands2022} and Lou et al.~\cite{Lou2022} further show that virtual
influencer reception involves parasocial bonds, perceived authenticity, and
appearance-based evaluation. Looi et al.~\cite{Looi2025} analyse Instagram
comments on Lil Miquela and find polarised reactions associated with identity
ambiguity, and Yan et al.~\cite{Yan2024} show that the design realism of VIs
shapes emotional attachment. These studies motivate our focus on how audiences
organise discourse around artificial identity, appearance, and emotional response.

\subsection{Computational analysis of social-media discourse}

Computational social science often studies online discourse through sentiment,
topic, stance, engagement, and language-style indicators. Such signals are useful
for large-scale analysis, but they are often treated independently. In this paper,
we combine sentiment labels, topic tags, and Big Five psycholinguistic style cues
within a single symbolic context. Big Five cues inferred from text have a long
tradition in computational psycholinguistics~\cite{Mairesse2007}; however, we
use them only as aggregate textual style indicators, not as psychological
measurements of individual commenters. This distinction is important because the
unit of analysis is the \emph{organisation of discourse}, not the diagnosis of
individual users.

\subsection{Symbolic and semantic interpretability}

Formal Concept Analysis~(FCA) provides a lattice-theoretic framework for
identifying closed co-occurrence structures in binary object--attribute
relations~\cite{GanterWille1999}. Prior work links FCA to frequent closed
itemset mining~\cite{Stumme2002}, efficient closed-set discovery~\cite{Pasquier1999},
and knowledge processing applications~\cite{Poelmans2013}. Association rule
mining~\cite{Agrawal1994} provides complementary summaries of directional
co-occurrence regularities through support, confidence, and lift. Our contribution
is to connect these symbolic structures to semantic embedding geometry: formal
concepts are rendered as natural-language descriptions, embedded using MiniLM,
and compared across HI and VI audiences. This symbolic--semantic combination
supports interpretable analysis of social discourse by preserving formal
co-occurrence structure while enabling semantic comparison of discourse regimes.

\section{Data}

\subsection{Dataset and influencer pairing}

We collect top-level YouTube comments via the YouTube Data API~(v3)
from three VI--HI pairs matched on content niche, subscriber scale,
and posting frequency.
We retain English-language comments (\texttt{langdetect}),
remove duplicates and empty strings, and strip HTML.

\begin{table}[!t]
\centering
\small
\setlength{\tabcolsep}{4pt}
\renewcommand{\arraystretch}{1.15}
\begin{tabular}{l l l l}
\toprule
Virtual Influencer & Subscribers & Human Counterpart & Niche \\
\midrule
Lil Miquela  & $\sim$271k & Samantha Nicole & Fashion / lifestyle \\
APOKI        & $\sim$329k & YOUNG POSSE     & K-pop / performance \\
Milla Sofia  & $\sim$28k  & Lydia Stoner    & Fashion lookbooks \\
\bottomrule
\end{tabular}
\caption{Selected VI--HI influencer pairs. HI counterparts are chosen by
automated API search within a $\pm$10\% subscriber window, then qualitatively
aligned on content niche.}
\label{tab:influencers}
\end{table}

Table~\ref{tab:influencers} summarises the pairings.
The final dataset contains \textbf{29,327 VI comments} and
\textbf{40,171 HI comments}. In total, the dataset comprises \textbf{69,498} comments across both conditions. To mitigate confounding, we match each HI within a narrow subscriber window and by niche/content format; nevertheless, residual differences (e.g., audience demographics and creator style) may remain, so our claims are framed as \emph{pair-matched structural contrasts} rather than causal effects of virtuality.
\subsection{Comment enrichment}

Each comment is enriched with three attribute families:

\paragraph{Sentiment.}
Three-class (positive/neutral/negative) classification via a
fine-tuned RoBERTa classifier.

\paragraph{Big Five personality cues.}
Continuous scores (Openness, Conscientiousness, Extraversion,
Agreeableness, Neuroticism) inferred via a transformer-based
regressor~\cite{Mairesse2007}.
Discretised to \attr{high}/\attr{low} binary attributes at the group-level
mean; ties are rare, and when they occur, we break them deterministically
by assigning greater-than-or-equal values to \attr{high}.

We treat inferred Big Five scores as \emph{psycholinguistic style proxies} extracted from text, not as ground-truth personality measurements of the commenters, and we use them only for \emph{comparative structural analysis} under an identical pipeline across conditions.
\paragraph{Topics.}
Twelve-class taxonomy assigned by Gemini Flash-Lite via zero-shot
NLI: \attr{positivity}, \attr{appearance}, \attr{authenticity\_critique},
\attr{artificial\_identity},\\ \attr{parasocial}, \attr{brand\_ads},
\attr{criticism}, \attr{humor}, \attr{performance},
\attr{mental\_health},\\ \attr{body\_image}, \attr{social\_comparison}.

The full attribute vocabulary is $|M|=25$ binary attributes, consisting of 3 sentiment attributes, 12 topic attributes, and 10 personality-bin attributes.
Table~\ref{tab:attr_prev} reports \emph{marginal} attribute prevalence (how often each signal appears in isolation).
In Section~\ref{sec:method_fca}, we quantify the resulting \emph{concept space} under FCA iceberg filtering, capturing how signals \emph{co-occur} structurally rather than individually.
\begin{table}[!t]
\centering

\setlength{\tabcolsep}{5pt}
\renewcommand{\arraystretch}{1.15}
\begin{tabular}{l c c c}
\toprule
Attribute & HI & VI & $\Delta$ (VI$-$HI) \\
\midrule
\attr{topic\_artificial\_identity} & 0.005 & 0.184 & $+$0.179 \\
\attr{sentiment\_Negative}         & 0.067 & 0.125 & $+$0.058 \\
\attr{topic\_authenticity\_critique} & 0.037 & 0.062 & $+$0.025 \\
\attr{topic\_appearance}           & 0.159 & 0.170 & $+$0.011 \\
\attr{topic\_social\_comparison}   & 0.007 & 0.007 & $\phantom{+}$0.000 \\
\attr{sentiment\_Positive}         & 0.766 & 0.669 & $-$0.097 \\
\attr{topic\_positivity}           & 0.677 & 0.472 & $-$0.205 \\
\bottomrule
\end{tabular}
\caption{Attribute prevalence (selected attributes), sorted by $\Delta$.
The near-equal appearance prevalence (0.011 gap) contrasts sharply with
its structural asymmetry in rule mining (Section~\ref{sec:rq2}).}
\label{tab:attr_prev}

\end{table}

\section{Methodology}
\subsection{FCA preliminaries: from formal contexts to rules}
\label{sec:fca_preliminaries}

Formal Concept Analysis represents data as a binary relation between
objects and attributes. A \emph{formal context} is a triple
$\mathbb{K}=(G,M,I)$, where $G$ is a set of objects, $M$ is a set of
attributes, and $I \subseteq G \times M$ indicates which object has which
attribute. In our setting, objects are either comments or weekly aggregates,
and attributes are binary discourse signals such as sentiment labels, topic
tags, and discretised psycholinguistic cues.

For $A \subseteq G$ and $B \subseteq M$, FCA defines two derivation operators:
\begin{align}
A' &= \{m \in M \mid \forall g \in A,\ (g,m)\in I\},\\
B' &= \{g \in G \mid \forall m \in B,\ (g,m)\in I\}.
\end{align}
Thus, $A'$ is the set of attributes shared by all objects in $A$, and $B'$ is
the set of objects sharing all attributes in $B$. A \emph{formal concept} is a
pair $(A,B)$ such that $A'=B$ and $B'=A$. The set $A$ is called the
\emph{extent}, and the set $B$ is called the \emph{intent}. Intuitively, a
formal concept is a maximal closed co-occurrence pattern: no object can be
added without losing an attribute, and no attribute can be added without losing
an object.

Table~\ref{tab:toy_context} gives a small illustrative context. Here, five
comments are described using four binary discourse attributes:
positive sentiment ($P$), appearance topic ($A$), high agreeableness ($G$),
and high neuroticism ($N$).

\begin{table}[t]
\centering
\small
\setlength{\tabcolsep}{8pt}
\renewcommand{\arraystretch}{1.15}
\begin{tabular}{lcccc}
\toprule
Object & $P$ & $A$ & $G$ & $N$ \\
\midrule
$c_1$ & \checkmark & \checkmark & \checkmark &  \\
$c_2$ & \checkmark & \checkmark & \checkmark &  \\
$c_3$ &  & \checkmark &  & \checkmark \\
$c_4$ & \checkmark &  & \checkmark &  \\
$c_5$ & \checkmark & \checkmark & \checkmark & \checkmark \\
\bottomrule
\end{tabular}
\caption{Toy formal context illustrating FCA. Objects are comments and
attributes are binary discourse signals: $P$ = positive sentiment,
$A$ = appearance topic, $G$ = high agreeableness, and $N$ = high neuroticism.}
\label{tab:toy_context}
\end{table}

From this context, FCA identifies closed attribute sets such as
$\{P,G\}$, $\{A,N\}$, and $\{P,A,G\}$. For instance, the attribute set
$\{A,G\}$ is not closed, because all comments that have both $A$ and $G$
also have $P$. Its closure is therefore $\{P,A,G\}$. This closed pattern
corresponds to the formal concept:
\[
(\{c_1,c_2,c_5\},\{P,A,G\}).
\]
In discourse terms, this concept represents a group of comments where
appearance-related discourse co-occurs with positive sentiment and high
agreeableness.

The formal concepts are partially ordered by extent inclusion:
\[
(A_1,B_1) \leq (A_2,B_2) \iff A_1 \subseteq A_2,
\]
or equivalently by reverse intent inclusion:
\[
B_2 \subseteq B_1.
\]
This ordering forms a concept lattice. Figure~\ref{fig:toy_lattice} shows the
lattice induced by the toy context. Moving downward in the lattice corresponds
to adding attributes and obtaining more specific discourse patterns.

\begin{figure}[t]
\centering
\begin{tikzpicture}[
    node distance=1.2cm and 1.6cm,
    every node/.style={draw, rounded corners, align=center, font=\scriptsize,
    inner sep=3pt},
    edge/.style={-}
]

\node (top) {$\emptyset$\\$\{c_1,c_2,c_3,c_4,c_5\}$};

\node[below left=of top] (A) {$\{A\}$\\$\{c_1,c_2,c_3,c_5\}$};
\node[below right=of top] (PG) {$\{P,G\}$\\$\{c_1,c_2,c_4,c_5\}$};

\node[below left=of A] (AN) {$\{A,N\}$\\$\{c_3,c_5\}$};
\node[below right=of A] (PAG) {$\{P,A,G\}$\\$\{c_1,c_2,c_5\}$};

\node[below=of PAG] (bottom) {$\{P,A,G,N\}$\\$\{c_5\}$};

\draw[edge] (top) -- (A);
\draw[edge] (top) -- (PG);
\draw[edge] (A) -- (AN);
\draw[edge] (A) -- (PAG);
\draw[edge] (PG) -- (PAG);
\draw[edge] (AN) -- (bottom);
\draw[edge] (PAG) -- (bottom);

\end{tikzpicture}
\caption{Toy concept lattice derived from Table~\ref{tab:toy_context}.
Each node shows a closed intent and its corresponding extent. More specific
concepts appear lower in the lattice.}
\label{fig:toy_lattice}
\end{figure}

Association rules are then derived from co-occurring attribute sets. For two
disjoint attribute sets $X,Y \subseteq M$, a rule $X \Rightarrow Y$ is evaluated
using support, confidence, and lift:
\begin{align}
\mathrm{supp}(X{\Rightarrow}Y) &= P(X \cup Y),\\
\mathrm{conf}(X{\Rightarrow}Y) &= P(Y \mid X),\\
\mathrm{lift}(X{\Rightarrow}Y) &= \frac{P(Y \mid X)}{P(Y)}.
\end{align}
For example, in Table~\ref{tab:toy_context}, the rule
\[
\{A,G\} \Rightarrow \{P\}
\]
has support $3/5$, confidence $1.0$, and lift $1/0.8=1.25$, because all
comments with both appearance and high agreeableness also have positive
sentiment. In the full study, the same logic is applied to the VI and HI
comment-level formal contexts, using support, confidence, and lift thresholds
to retain interpretable discourse rules.
Our methodology combines symbolic structure discovery and semantic geometry.
First, FCA extracts closed co-occurrence patterns from weekly and comment-level
formal contexts. Second, association rules summarise directional regularities
among sentiment, topic, and psycholinguistic attributes. Third, we render formal
concepts as natural-language descriptions, embed them using MiniLM, and compare
semantic geometry across HI and VI concept spaces. Finally, we perform targeted
topic-level analyses for psychologically salient domains.

Table~\ref{tab:analytical_views} summarises the role of each analytical layer.
The framework is designed so that each layer answers a different social-science
question: marginal analysis describes what is frequent, FCA and rules describe
what co-occurs, semantic geometry describes how closed concepts relate in
meaning space, and cluster labels provide human-readable interpretations of
computed discourse regions.

\begin{table}[!t]
\centering
\scriptsize
\setlength{\tabcolsep}{4pt}
\renewcommand{\arraystretch}{1.15}
\begin{tabular}{p{2.6cm}p{4.0cm}p{4.8cm}}
\toprule
Analytical view & Captures & Contribution to the study \\
\midrule
Marginal prevalence & Individual sentiment/topic/style frequencies & Establishes baseline differences and exposes cases where similar frequencies hide different structures. \\
FCA concepts & Closed co-occurrence bundles & Identifies stable symbolic discourse profiles that cannot be reduced to individual attributes. \\
Association rules & Directional co-occurrence regularities & Summarises interpretable discourse grammars through support, confidence, and lift. \\
MiniLM concept geometry & Semantic proximity among formal concepts & Tests whether symbolic structures correspond to coherent semantic regions. \\
Cluster labels & Human-readable discourse regimes & Interprets computed semantic regions without replacing the symbolic evidence. \\
\bottomrule
\end{tabular}
\caption{Analytical views in the symbolic--semantic discourse framework.}
\label{tab:analytical_views}
\end{table}

\begin{figure}[!t]
\centering
\includegraphics[width=0.76\linewidth]{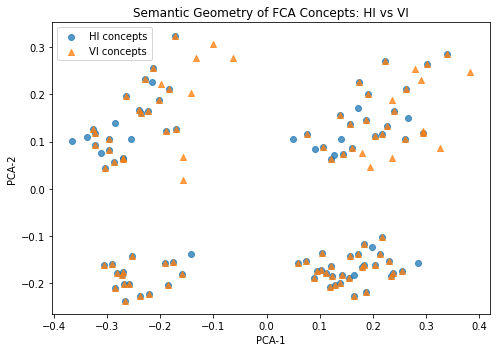}
\caption{Semantic geometry of FCA concepts for HI and VI audiences. Each point
represents a formal concept rendered as a natural-language description and
embedded using MiniLM. HI and VI concepts partially overlap, while VI concepts
occupy a slightly more dispersed semantic region.}
\label{fig:semantic_geometry_overlay}
\end{figure}

Figure~\ref{fig:semantic_geometry_overlay} previews the semantic concept space
used in the final stage of the analysis. HI and VI concepts partially overlap,
indicating shared audience response modes, while VI concepts occupy a slightly
broader region, suggesting greater discourse heterogeneity.

\subsection{Layer 1: Weekly FCA with support-based iceberg filtering}
\label{sec:method_fca}
\paragraph{Aggregation.}
Comments are grouped by calendar week.
Within each week, sentiment and topic are summarised by modal label;
Big Five traits are averaged across comments and then binarised
against the group-level weekly mean.
Each week becomes one binary object in the formal context.

\paragraph{Formal context and concept extraction.}
We build one binary formal context
$\mathbb{K}_c=(G_c, M, I_c)$ per condition $c \in \{\mathrm{HI},\mathrm{VI}\}$,
where $G_c$ is the set of weeks and $M$ is the shared attribute vocabulary.
FCA extracts
formal concepts;
a \emph{formal concept} $(A,B)$ satisfies $A=B'$ and $B=A'$, where
$B' = \{g \in G_c \mid \forall m \in B: (g,m)\in I_c\}$ and symmetrically.
We order concepts by $\subseteq$ on extents to form the concept
lattice $\mathcal{L}(\mathbb{K}_c)$~\cite{GanterWille1999}.

\paragraph{Iceberg filtering.}
FCA can generate a large number of concepts, many weakly supported
or short-lived.
We retain only those with:
(i)~\textbf{minimum temporal support $\geq$\,20\%}
(concept must appear in at least 20\% of all weeks), and
(ii)~\textbf{minimum intent size $\geq$\,3}
(concept must encode at least three distinct attributes).

These thresholds were selected empirically to balance interpretability
and coverage: lower supports produced large numbers of weak concepts,
while higher thresholds eliminated meaningful discourse patterns.

Table~\ref{tab:fca_ablation} shows
the dramatic compression achieved:
from 1,895 VI concepts to 10 filtered, and from 864 HI concepts to 24.
These thresholds balance coverage and interpretability and are used
for all subsequent weekly-level analyses. Unlike Table 2 (marginals), Table 3 quantifies the size of the closed co-occurrence space under FCA filtering.

\begin{table}[!t]
\centering
\tiny
\setlength{\tabcolsep}{5pt}
\renewcommand{\arraystretch}{1.10}
\begin{tabular}{c c c r r r r}
\toprule
Multilingual & Intent & Support & VI (raw) & VI (filtered) & HI (raw) & HI (filtered) \\
\midrule
No  & $>$3 & 0.20 & 1895 & \textbf{10}  & 864 & \textbf{24} \\
No  & $>$3 & 0.10 & 1895 & 114 & 864 & 152 \\
No  & $>$2 & 0.20 & 1895 & 54  & 864 & 80  \\
No  & $>$2 & 0.10 & 1895 & 241 & 864 & 264 \\
\midrule
Yes & $>$3 & 0.20 & 3980 & 11  & 1269 & 24  \\
Yes & $>$3 & 0.10 & 3980 & 124 & 1269 & 153 \\
Yes & $>$2 & 0.20 & 3980 & 57  & 1269 & 80  \\
Yes & $>$2 & 0.10 & 3980 & 288 & 1269 & 268 \\
\bottomrule
\end{tabular}
\caption{FCA concept counts under different support, intent-size, and
language-inclusion settings (Multilingual = Yes includes non-English comments).
Bold row shows the selected configuration (English-only, intent $>$3, support $\geq$0.20).}
\label{tab:fca_ablation}
\end{table}

\paragraph{Cross-group comparison.}
Concepts with identical intents across conditions are labelled
\emph{shared}; those unique to one condition are labelled
\emph{VI-only} or \emph{HI-only}.
We visualise attribute prevalence within each category using
horizontal bar charts.

%

\subsection{Layer 2: Comment-level association rule mining}

\paragraph{Context construction.}
We build a second formal context per condition where objects are
individual comments (not weeks) and attributes are the same 25
binary signals.
For comment-level rules, we discretise continuous personality traits at the comment level globally, whereas for weekly FCA we discretise weekly-averaged traits locally.

\paragraph{Rule generation.}
We use an approximate concept-based rule-mining procedure to generate candidate
association rules $X \Rightarrow Y$ ($X \cap Y = \emptyset$, $|X| \leq 3$),
derived from closed co-occurrence patterns in the formal context.
We then filter candidates by support, confidence, and lift:
\begin{align}
  \mathrm{supp}(X{\Rightarrow}Y) &= P(X\cup Y), &
  \mathrm{conf}(X{\Rightarrow}Y) &= P(Y \mid X), &
  \mathrm{lift}(X{\Rightarrow}Y) &= \frac{P(Y \mid X)}{P(Y)}.
\end{align}
Rules are retained if minsup\,$\geq$\,1\%, minconf\,$\geq$\,0.8,
and lift\,$>$\,1.2.

We report \emph{association rules} (statistical regularities quantified by support/confidence/lift), and do not interpret them as logical implications or compute an implication basis (e.g., Duquenne--Guigues).
\paragraph{Rule cluster analysis.}
We group retained rules by the dominant topic attribute in their
antecedent to identify \emph{premise clusters}---interpretable
discourse sub-grammars sharing a topic anchor.

\subsection{Semantic Geometry of Formal Concepts}
\label{sec:method_semantic_geometry}

To examine whether symbolic FCA structures correspond to meaningful semantic
regions, we transform each formal concept intent into a short natural-language
description. For example, the intent
\{\attr{topic\_appearance}, \attr{sentiment\_Positive},
\attr{Agreeableness\_high}, \attr{Conscientiousness\_low}\} is rendered as:
``Audience discourse focused on appearance, with positive sentiment, associated
with high agreeableness and low conscientiousness.''

We embed these concept descriptions using MiniLM and cluster the resulting
vectors using agglomerative clustering with cosine distance. This yields a
semantic concept space in which each point corresponds to a closed symbolic
discourse pattern. We then compare VI and HI concept spaces using two measures.
First, \emph{semantic dispersion} is computed as the average cosine distance
between each concept embedding and the group centroid. Higher dispersion
indicates that extracted discourse concepts occupy a broader semantic region.
Second, \emph{symbolic--semantic alignment} is computed by correlating FCA
symbolic distance, measured by Jaccard distance between concept intents, with
MiniLM semantic distance. A high positive correlation indicates that symbolic
proximity in the closed-set structure is reflected in semantic embedding space.

\subsection{Complementary topic-level analysis}

For topics that may fluctuate week-to-week and therefore resist
weekly FCA, we conduct a direct comment-level analysis
(without time aggregation) of four psychologically salient topics:
artificial identity, authenticity critique, body image, and mental health.
For each topic and each condition, we compute:
(i)~sentiment distribution (positive/neutral/negative proportions);
(ii)~average Big Five personality profiles, visualised as radar charts.

For clarity of presentation, we report (i) marginal prevalences (Table~\ref{tab:attr_prev}), (ii) iceberg-filtered concept-space sizes (Table~\ref{tab:fca_ablation}), and (iii) clustered rule sets under a robustness grid (Table~\ref{tab:grid}), aligning each result with RQ1--RQ3.

\subsection{Reliability, robustness, and interpretation safeguards}
\label{sec:method_safeguards}

Several design choices are used to make the comparison conservative. First, the
same annotation, binarisation, concept extraction, rule filtering, embedding, and
clustering pipeline is applied to VI and HI comments. Second, association rules
are evaluated under a support--confidence grid rather than a single threshold,
allowing us to check whether key findings, especially the VI-specific appearance
rules, persist across parameter settings. Third, semantic labels are assigned only
after formal concepts have been extracted and embedded; the labels are used as
interpretive summaries of computed clusters, not as evidence-generating steps.
Finally, all findings are reported as structural associations in pair-matched
observational data rather than causal effects of virtuality.

\section{Results}

\subsection{RQ1a: Weekly symbolic discourse profiles}
\label{sec:rq1}
\begin{figure}[h]
\centering
\begin{minipage}[t]{0.49\textwidth}
  \centering
  \includegraphics[width=\linewidth]{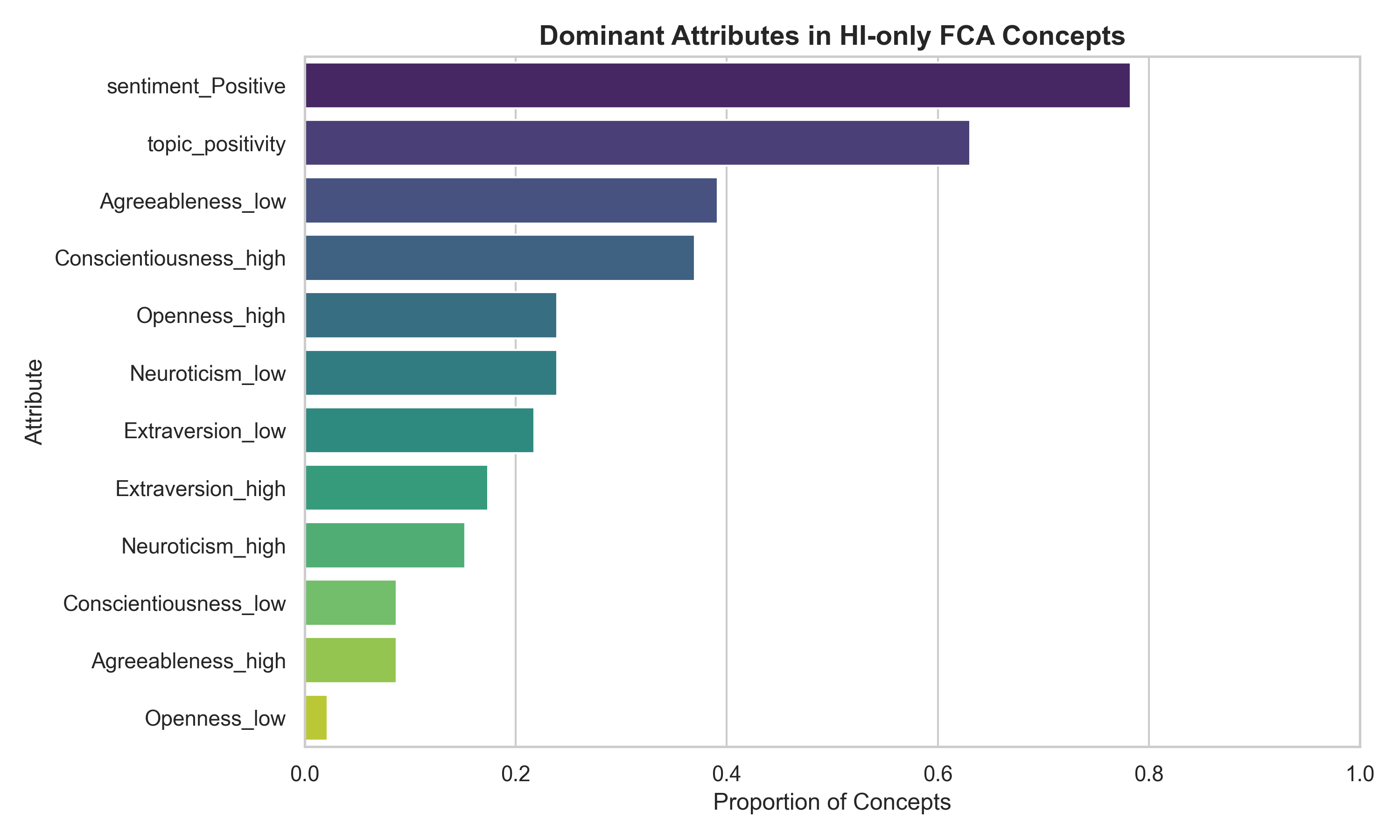}
  \vspace{-2mm}
  \subcaption{\textbf{} HI-only stable concepts}
\end{minipage}\hfill
\begin{minipage}[t]{0.49\textwidth}
  \centering
  \includegraphics[width=\linewidth]{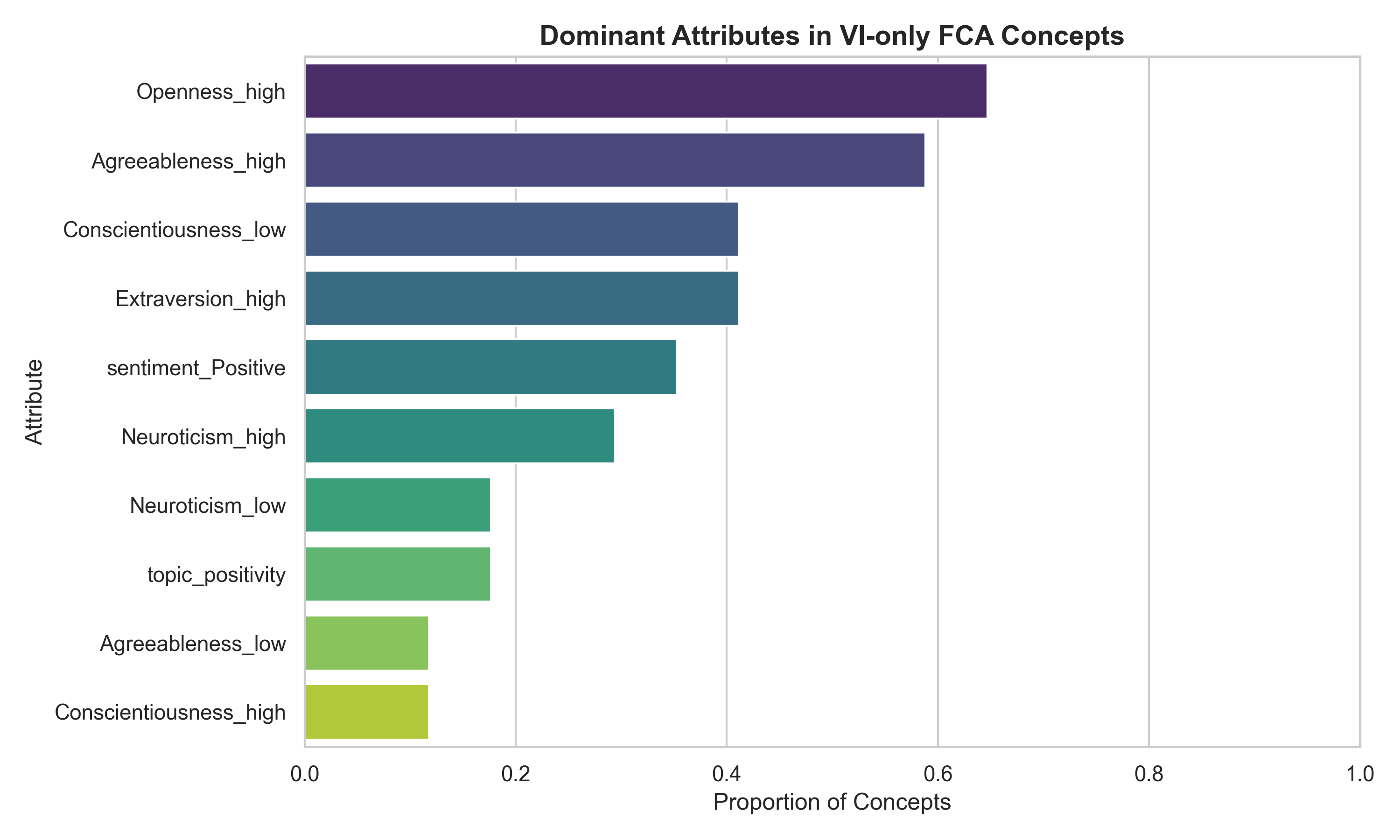}
  \vspace{-2mm}
  \subcaption{\textbf{} VI-only stable concepts}
\end{minipage}

\caption{Dominant attributes in group-specific (HI-only vs.\ VI-only) stable FCA concepts.
HI-only concepts are sentiment-centred (positive sentiment, topic positivity, low neuroticism),
whereas VI-only concepts are personality-centred (high openness, high agreeableness, low conscientiousness).}
\label{fig:hi_vi_only_fca}
\end{figure}
\

Figure~\ref{fig:hi_vi_only_fca} compares HI-only
and VI-only stable concept attribute distributions.

\textbf{HI-only concepts} are sentiment-dominated:
positive sentiment and topic positivity are the most frequent attributes,
alongside low neuroticism and high conscientiousness.
This indicates that HI discourse is organised around explicit
affective evaluation in a psychologically stable configuration.

\textbf{VI-only concepts} are personality-dominated:
high openness is the most frequent attribute,
followed by high agreeableness, low conscientiousness, and high extraversion.
Positive sentiment and topic positivity are less prominent.
High neuroticism appears more frequently in VI-only concepts,
indicating greater emotional intensity.
VI discourse appears exploratory and expressiveness-centred rather
than evaluation-centred.

This structural distinction---sentiment-centred HI vs.\ personality-centred VI---constitutes a qualitative difference not visible from Table~\ref{tab:attr_prev}.

\subsection{RQ1b: Comment-level rule grammar}
\label{sec:rq2}

\paragraph{Overall density.}
Table~\ref{tab:rule_counts} shows that VI yields 51 rules and HI yields 8
under identical thresholds---a 6.4-fold difference.
Table~\ref{tab:rule_summary} confirms higher average lift in VI (1.32 vs.\ 1.23)
and much higher maximum support (0.42 vs.\ 0.09).

\begin{table}[t!]
\centering
\setlength{\tabcolsep}{6pt}\renewcommand{\arraystretch}{1.15}
\begin{tabular}{l c c}
\toprule
Configuration & \#HI rules & \#VI rules \\
\midrule
minsup\,=\,1\%, minconf\,=\,0.8, lift\,$>$\,1.2 & 8 & 51 \\
\bottomrule
\end{tabular}
\caption{Rule-set sizes under the approximate concept-based rule-mining procedure, after support, confidence, and lift filtering.}
\label{tab:rule_counts}
\end{table}

\begin{table}[t!]
\centering
\setlength{\tabcolsep}{6pt}\renewcommand{\arraystretch}{1.15}
\begin{tabular}{l c c c}
\toprule
Metric & Mean & Median & Max \\
\midrule
\multicolumn{4}{l}{\textbf{HI rules (8 total)}} \\
Support    & 0.062 & 0.065 & 0.090 \\
Confidence & 0.92  & 0.94  & 0.96  \\
Lift       & 1.23  & 1.23  & 1.25  \\
\midrule
\multicolumn{4}{l}{\textbf{VI rules (51 total)}} \\
Support    & 0.146 & 0.080 & 0.420 \\
Confidence & 0.88  & 0.88  & 0.96  \\
Lift       & 1.32  & 1.32  & 1.43  \\
\bottomrule
\end{tabular}
\caption{Rule summary statistics (minsup\,=\,1\%, minconf\,=\,0.8, lift\,$>$\,1.2).}
\label{tab:rule_summary}
\end{table}

\paragraph{HI rule structure.}
All 8 HI rules share a single premise cluster:
\attr{topic\_positivity} $\wedge$ \attr{Neuroticism\_low} predicting
\attr{sentiment\_Positive} (conf 0.92--0.96, lift 1.20--1.25).
Low neuroticism is the constant structural driver;
secondary traits (agreeableness, extraversion, conscientiousness)
modulate but do not determine the association.
This indicates a \emph{stability-centred} discourse architecture:
positive sentiment in HI is anchored in psychological regulation.

\paragraph{VI rule structure.}
VI rules form three distinct premise clusters:
\begin{enumerate}[leftmargin=*]
\item \textbf{Positivity-framing} (40 rules): \attr{topic\_positivity} is
      the topic anchor; all Big Five configurations appear as secondary
      antecedents, including high neuroticism
      (conf\,=\,0.92, lift\,=\,1.37)---absent from HI.
      Positivity in VI is \emph{expressiveness-centred}: it persists
      even under emotional intensity.
\item \textbf{Appearance-discourse} (7 rules): \attr{topic\_appearance}
      is the topic anchor, predicting \attr{sentiment\_Positive}
      (conf 0.81--0.88, lift 1.21--1.32).
      This cluster is \textbf{entirely absent from HI}
      despite near-equal appearance prevalence (0.170 vs.\ 0.159,
      Table~\ref{tab:attr_prev})---the central marginal-versus-structure finding.
      Importantly, no HI rules contain \attr{topic\_appearance} under the full grid
      of tested thresholds (minsup $\in \{0.005, 0.01, 0.02\}$, minconf $\in \{0.8, 0.9\}$); 
      appearance rules remain VI-specific even when varying minsup and minconf.
\item \textbf{Personality-only} (4 rules): no topic attribute in antecedent;
      \attr{Agreeableness\_high} $\wedge$ \attr{Neuroticism\_low}
      alone predicts \attr{sentiment\_Positive}
      (conf 0.81--0.85, lift 1.21--1.27).
\end{enumerate}

Tables~\ref{tab:hi_rules}--\ref{tab:vi_rules_app} list the full HI rule set
and the top VI rules by cluster.

\begin{table}[h]
\centering\scriptsize
\setlength{\tabcolsep}{4pt}\renewcommand{\arraystretch}{1.15}
\begin{tabular}{p{6.0cm} p{2.4cm} c c c}
\toprule
Premise $X$ & Consequent $Y$ & supp & conf & lift \\
\midrule
\attr{Agreeableness\_high} $\wedge$ \attr{Neuroticism\_low} $\wedge$ \attr{topic\_positivity}
  & \attr{sentiment\_Positive} & 0.060 & 0.96 & 1.25 \\
\attr{Extraversion\_high} $\wedge$ \attr{Neuroticism\_low} $\wedge$ \attr{topic\_positivity}
  & \attr{sentiment\_Positive} & 0.080 & 0.95 & 1.24 \\
\attr{Conscientiousness\_low} $\wedge$ \attr{Neuroticism\_low} $\wedge$ \attr{topic\_positivity}
  & \attr{sentiment\_Positive} & 0.070 & 0.95 & 1.24 \\
\attr{Neuroticism\_low} $\wedge$ \attr{topic\_positivity}
  & \attr{sentiment\_Positive} & 0.090 & 0.94 & 1.23 \\
\attr{Neuroticism\_low} $\wedge$ \attr{Openness\_high} $\wedge$ \attr{topic\_positivity}
  & \attr{sentiment\_Positive} & 0.090 & 0.94 & 1.23 \\
\attr{Agreeableness\_high} $\wedge$ \attr{Neuroticism\_low} $\wedge$ \attr{sentiment\_Positive}
  & \attr{topic\_positivity}   & 0.060 & 0.82 & 1.21 \\
\attr{Agreeableness\_low} $\wedge$ \attr{Neuroticism\_low} $\wedge$ \attr{topic\_positivity}
  & \attr{sentiment\_Positive} & 0.030 & 0.92 & 1.20 \\
\attr{Conscientiousness\_high} $\wedge$ \attr{Neuroticism\_low} $\wedge$ \attr{topic\_positivity}
  & \attr{sentiment\_Positive} & 0.020 & 0.92 & 1.20 \\
\bottomrule
\end{tabular}
\caption{All 8 HI rules. Single cluster: \attr{topic\_positivity} $\wedge$
\attr{Neuroticism\_low} anchors all rules.}
\label{tab:hi_rules}
\end{table}

\begin{table}[h]
\centering\scriptsize
\setlength{\tabcolsep}{4pt}\renewcommand{\arraystretch}{1.15}
\begin{tabular}{p{6.0cm} p{2.4cm} c c c}
\toprule
Premise $X$ & Consequent $Y$ & supp & conf & lift \\
\midrule
\multicolumn{5}{l}{\textit{Positivity-framing cluster (top rules by lift)}} \\
\attr{Agreeableness\_high} $\wedge$ \attr{Neuroticism\_low} $\wedge$ \attr{topic\_positivity}
  & \attr{sentiment\_Positive} & 0.050 & 0.96 & 1.43 \\
\attr{Neuroticism\_low} $\wedge$ \attr{topic\_positivity}
  & \attr{sentiment\_Positive} & 0.080 & 0.94 & 1.40 \\
\attr{Agreeableness\_high} $\wedge$ \attr{Neuroticism\_high} $\wedge$ \attr{topic\_positivity}
  & \attr{sentiment\_Positive} & 0.200 & 0.92 & 1.37 \\
\attr{topic\_positivity}
  & \attr{sentiment\_Positive} & 0.420 & 0.89 & 1.33 \\
  \midrule
\multicolumn{5}{l}{\textit{Appearance-discourse cluster (absent from HI)}} \\
\attr{Agreeableness\_high} $\wedge$ \attr{Conscientiousness\_low} $\wedge$ \attr{topic\_appearance}
  & \attr{sentiment\_Positive} & 0.070 & 0.88 & 1.32 \\
\attr{Agreeableness\_high} $\wedge$ \attr{Neuroticism\_high} $\wedge$ \attr{topic\_appearance}
  & \attr{sentiment\_Positive} & 0.070 & 0.87 & 1.30 \\
\attr{Agreeableness\_high} $\wedge$ \attr{topic\_appearance}
  & \attr{sentiment\_Positive} & 0.090 & 0.85 & 1.27 \\
\bottomrule
\end{tabular}
\caption{Selected VI rules. Top positivity-framing rules include
Neuroticism\_high---absent from HI. Appearance-discourse rules have
no HI counterpart despite equal marginal prevalence.}
\label{tab:vi_rules_app}
\end{table}

\subsection{Rule Extraction Robustness}
To confirm that the 6.4-fold difference in rule counts (and the specific presence of appearance rules in VI) is not an artifact of the chosen thresholds (minsup=1\%, minconf=0.8), we tested a grid of parameters. Table~\ref{tab:grid} demonstrates that VI consistently generates substantially more rules than HI, and that \attr{topic\_appearance} rules remain exclusively found in VI discourse across all tested configurations.

\begin{table}[h]
\centering
\setlength{\tabcolsep}{6pt}\renewcommand{\arraystretch}{1.15}
\begin{tabular}{c c c c}
\toprule
minsup & minconf & \#HI rules & \#VI rules \\
\midrule
0.005 & 0.8 & 22 & 114 \\
0.005 & 0.9 & 14 & 68 \\
0.010 & 0.8 & 8  & 51 \\
0.010 & 0.9 & 6  & 29 \\
0.020 & 0.8 & 3  & 18 \\
\bottomrule
\end{tabular}
\caption{Rule counts across extraction thresholds (lift\,$>$\,1.2) for selected configurations.}
\label{tab:grid}
\end{table}

\subsection{RQ2: Symbolic--semantic geometry of FCA concepts}
\label{sec:rq2_semantic_geometry}

To test whether FCA concepts correspond to meaningful semantic regions, we
embedded natural-language renderings of the top closed concepts using MiniLM and
compared their semantic distances with symbolic distances between concept
intents. Table~\ref{tab:symbolic_semantic_geometry} reports the resulting
alignment and dispersion measures. Both HI and VI concept spaces show strong
symbolic--semantic alignment: Spearman correlations between FCA symbolic
distance and MiniLM semantic distance are $\rho=0.614$ for HI and $\rho=0.640$
for VI ($p<0.001$ in both cases). This indicates that closed-set structure is
not merely a symbolic artefact: concepts that are close in FCA intent space also
tend to be close in semantic embedding space.

At the same time, VI concepts exhibit higher semantic dispersion than HI
concepts (0.0882 vs.\ 0.0765), suggesting that VI discourse spans a broader and
more heterogeneous semantic region. This refines the rule-mining result: VI
discourse is not simply less coherent; rather, it remains symbolically aligned
while covering a more differentiated semantic space.

\begin{table}[!t]
\centering
\small
\setlength{\tabcolsep}{6pt}
\renewcommand{\arraystretch}{1.15}
\begin{tabular}{lccc}
\toprule
Group & Symbolic--Semantic $\rho$ & $p$-value & Semantic Dispersion \\
\midrule
HI & 0.614 & $<0.001$ & 0.0765 \\
VI & 0.640 & $<0.001$ & 0.0882 \\
\bottomrule
\end{tabular}
\caption{Symbolic--semantic geometry of FCA concepts. $\rho$ is the Spearman
correlation between FCA symbolic distance, computed from concept intent overlap,
and MiniLM semantic distance, computed from concept-sentence embeddings.
Semantic dispersion is the mean cosine distance from each concept embedding to
the group centroid.}
\label{tab:symbolic_semantic_geometry}
\end{table}

\begin{figure}[!t]
\centering
\begin{minipage}{0.48\textwidth}
\centering
\includegraphics[width=\linewidth]{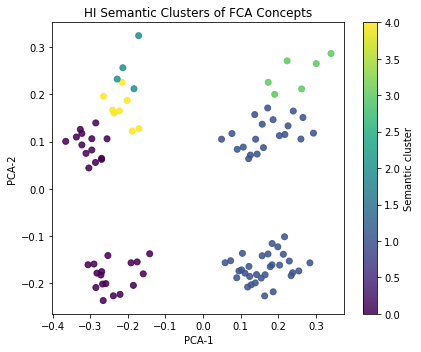}
\subcaption{HI semantic clusters}
\end{minipage}
\hfill
\begin{minipage}{0.48\textwidth}
\centering
\includegraphics[width=\linewidth]{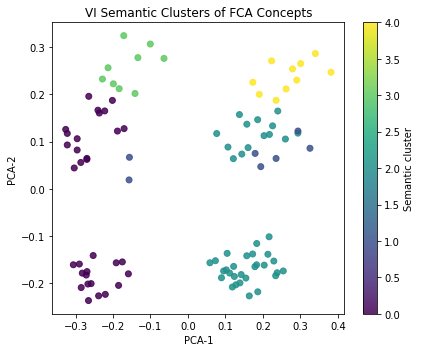}
\subcaption{VI semantic clusters}
\end{minipage}
\caption{Semantic clustering of FCA concepts. Each point corresponds to a
closed concept embedded using MiniLM and coloured by semantic cluster. HI
clusters are mainly organised around positivity and appearance-related
engagement, whereas VI clusters include a distinct artificial-identity region
and multiple appearance-related modes.}
\label{fig:semantic_clusters}
\end{figure}

Figure~\ref{fig:semantic_clusters} shows the semantic clusters of HI and VI
concepts. HI clusters are mainly organised around positivity and
appearance-related engagement. VI clusters include a distinct artificial-identity
region in addition to multiple appearance-related modes, indicating that
audience reactions to virtual influencers are semantically more differentiated
around identity, appearance, and emotionally reactive engagement.

\begin{figure}[!t]
\centering
\begin{minipage}{0.48\textwidth}
\centering
\includegraphics[width=\linewidth]{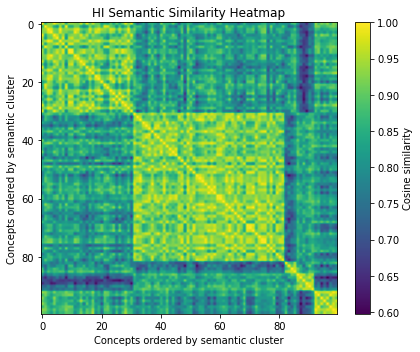}
\subcaption{HI semantic similarity}
\end{minipage}
\hfill
\begin{minipage}{0.48\textwidth}
\centering
\includegraphics[width=\linewidth]{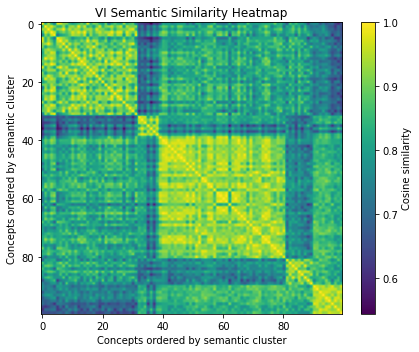}
\subcaption{VI semantic similarity}
\end{minipage}
\caption{Semantic similarity heatmaps of FCA concepts ordered by semantic
cluster. Block structure indicates coherent semantic regions among closed
discourse patterns. VI concepts show slightly higher semantic dispersion,
consistent with a broader range of audience response modes.}
\label{fig:semantic_heatmaps}
\end{figure}

The semantic similarity heatmaps in Figure~\ref{fig:semantic_heatmaps} show
clear block structure, confirming that concept clusters form coherent semantic
regions rather than arbitrary partitions. Compared with HI, the VI heatmap shows
slightly stronger differentiation across regions, consistent with the higher
semantic dispersion observed for VI concepts.

Table~\ref{tab:semantic_cluster_labels} summarises the cluster-level
interpretation. The labels are used only as human-readable summaries of clusters
computed from FCA-derived concept embeddings; the underlying evidence remains
the symbolic closed-set structure and the embedding geometry.

\begin{table}[!t]
\centering
\scriptsize
\setlength{\tabcolsep}{4pt}
\renewcommand{\arraystretch}{1.15}
\begin{tabular}{lll}
\toprule
Group & Cluster & Semantic label \\
\midrule
HI & C0 & Emotionally tense positive discourse \\
HI & C1 & High-arousal expressive positivity \\
HI & C2 & Tense appearance evaluation \\
HI & C3 & Expressive appearance admiration \\
HI & C4 & General positivity grammar \\
\midrule
VI & C0 & Emotionally reactive positive framing \\
VI & C1 & Artificial identity scrutiny \\
VI & C2 & High-arousal expressive engagement \\
VI & C3 & Warm but tense appearance evaluation \\
VI & C4 & Expressive aesthetic admiration \\
\bottomrule
\end{tabular}
\caption{Semantic labels assigned to MiniLM clusters of formal concepts. Labels
summarise clusters of closed itemsets after rendering concept intents as
natural-language descriptions.}
\label{tab:semantic_cluster_labels}
\end{table}

\subsection{RQ3: Topic-level sentiment and psycholinguistic patterns}
\label{sec:rq3}

\paragraph{Artificial identity.}
\begin{figure}[t!]
\centering
\begin{minipage}{0.4\textwidth}
\includegraphics[width=\linewidth]{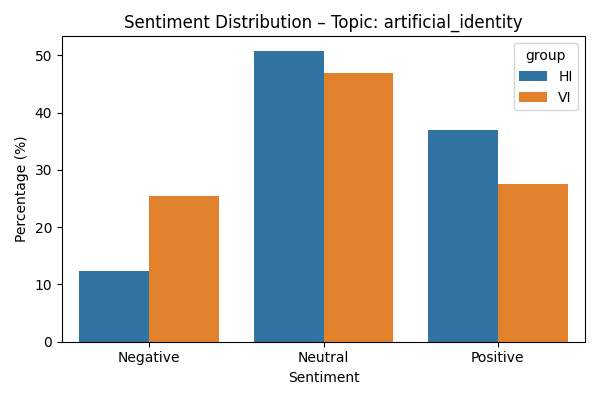}
\caption{Sentiment distribution: \textit{artificial identity}.
VI shows $\approx$25\% negative vs.\ $\approx$12\% for HI.}
\label{fig:ai_sent}
\end{minipage}\hfill
\begin{minipage}{0.4\textwidth}
\includegraphics[width=\linewidth]{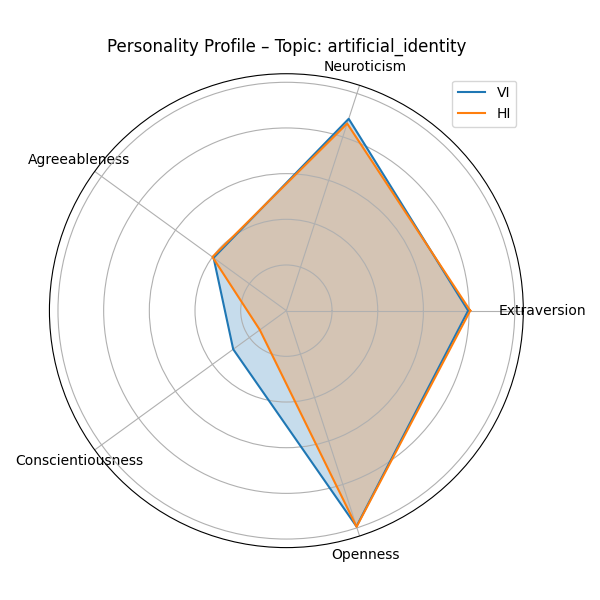}
\caption{Big Five profile: \textit{artificial identity}.
VI commenters score lower on Agreeableness and Conscientiousness.}
\label{fig:ai_traits}
\end{minipage}
\end{figure}

Both groups are predominantly neutral on this topic
(47.2\% VI, 50.8\% HI).
However, VI comments show substantially more negative sentiment
(24.8\% vs.\ 12.3\%) and less positive sentiment
(28.0\% vs.\ 36.9\%).
The personality profile (Fig.~\ref{fig:ai_traits}) shows comparable
openness and neuroticism, but VI commenters score lower on
agreeableness and conscientiousness, suggesting more evaluative and
critical discourse framing in virtual contexts.

\paragraph{Authenticity critique.}
\begin{figure}[h]
\centering
\begin{minipage}{0.4\textwidth}
\includegraphics[width=\linewidth]{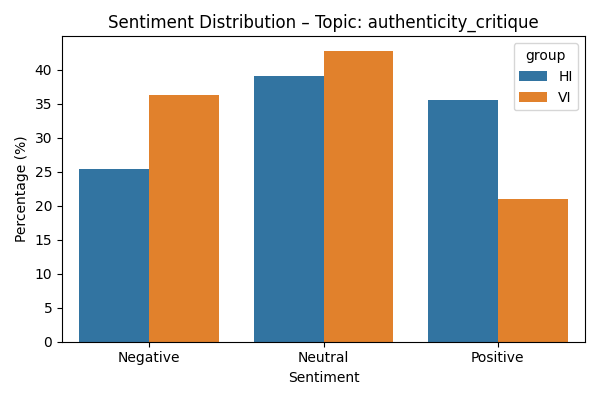}
\caption{Sentiment: \textit{authenticity critique}.
VI: $\approx$36\% negative; HI: $\approx$25\%.}
\label{fig:auth_sent}
\end{minipage}\hfill
\begin{minipage}{0.4\textwidth}
\includegraphics[width=\linewidth]{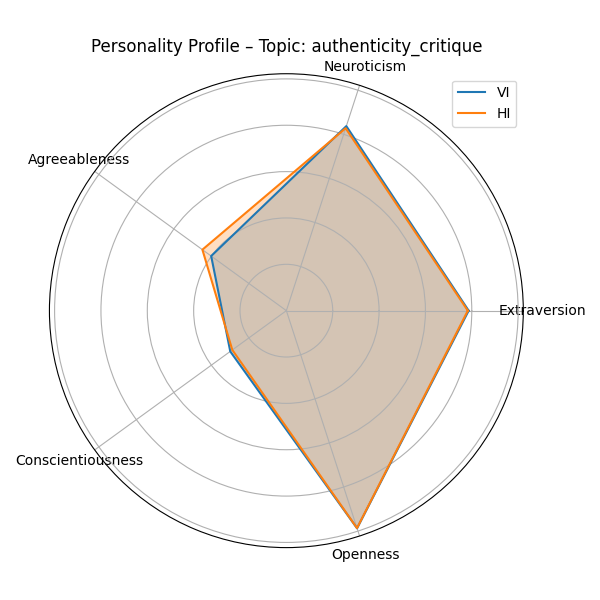}
\caption{Big Five profile: \textit{authenticity critique}.
Personality profiles are similar; VI shows slightly lower agreeableness.}
\label{fig:auth_traits}
\end{minipage}
\end{figure}

Both groups are predominantly neutral (41.6\% VI, 39.4\% HI),
but VI threads exhibit more negative sentiment (36.1\% vs.\ 24.8\%)
and less positive (22.3\% vs.\ 35.8\%).
Personality profiles are broadly similar (Fig.~\ref{fig:auth_traits}),
with high openness and neuroticism in both groups, suggesting
reflective and emotionally sensitive engagement regardless of type,
but the emotional direction diverges: VI audiences respond
more sceptically to authenticity violations.

\paragraph{Body image.}
\begin{figure}[t!]
\centering
\begin{minipage}{0.4\textwidth}
\includegraphics[width=\linewidth]{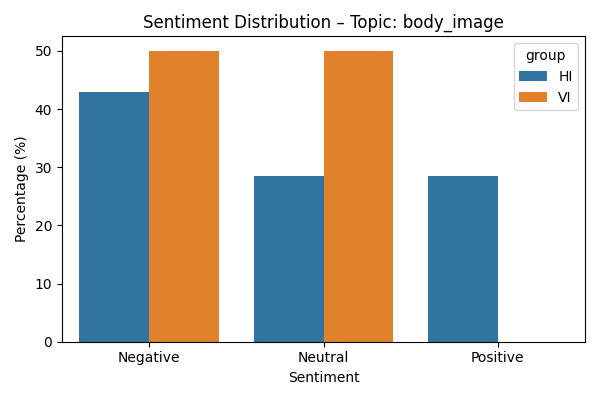}
\caption{Sentiment: \textit{body image}.
VI is overwhelmingly negative/neutral; HI retains $\approx$28\% positive.}
\label{fig:body_sent}
\end{minipage}\hfill
\begin{minipage}{0.4\textwidth}
\includegraphics[width=\linewidth]{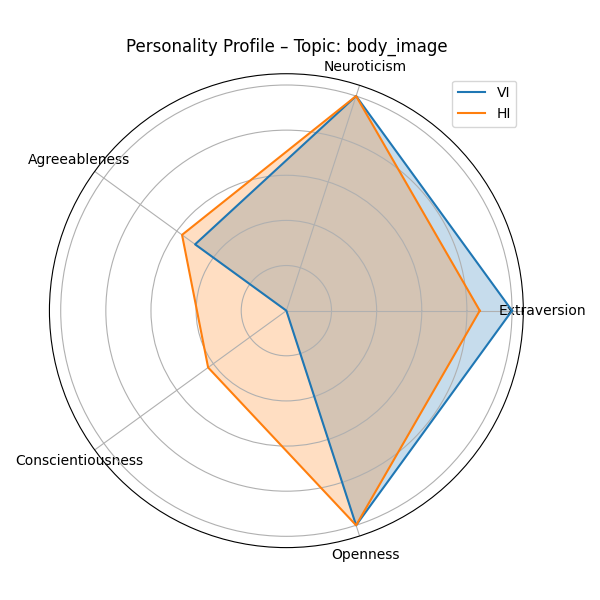}
\caption{Big Five profile: \textit{body image}.
HI shows higher agreeableness and conscientiousness.}
\label{fig:body_traits}
\end{minipage}
\end{figure}

VI body-image comments are overwhelmingly negative or neutral
with virtually no positive sentiment, whereas HI threads include
28.4\% positive (Fig.~\ref{fig:body_sent}).
HI personality profiles show higher agreeableness and conscientiousness
(Fig.~\ref{fig:body_traits}), indicating more supportive, structured
engagement; VI profiles show higher extraversion and lower conscientiousness,
suggesting more expressive but less moderated discourse.
\paragraph{Mental health.}
\begin{figure}[t!]
\centering
\begin{minipage}{0.4\textwidth}
\includegraphics[width=\linewidth]{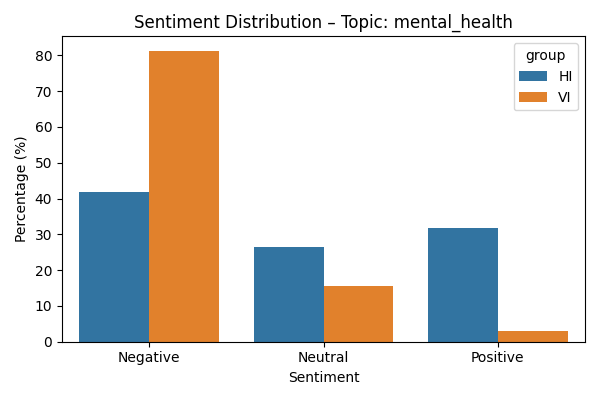}
\caption{Sentiment: \textit{mental health}.
VI: $\approx$81\% negative. HI: $\approx$42\% negative, $\approx$31\% positive.}
\label{fig:mh_sent}
\end{minipage}\hfill
\begin{minipage}{0.4\textwidth}
\includegraphics[width=\linewidth]{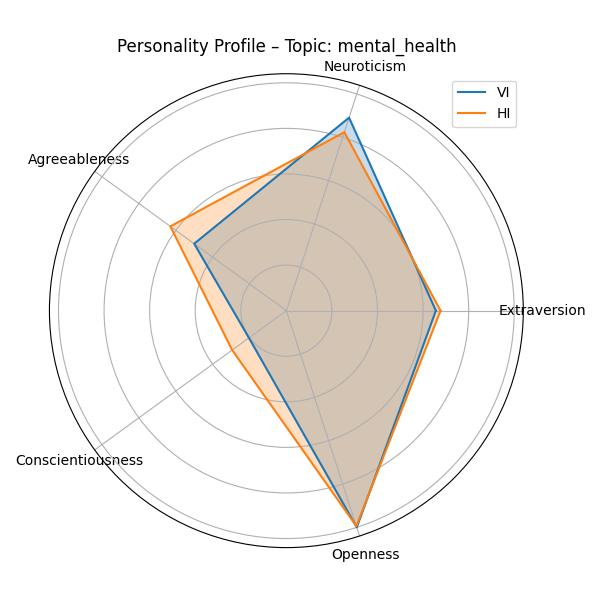}
\caption{Big Five profile: \textit{mental health}.
VI shows higher neuroticism and lower conscientiousness.}
\label{fig:mh_traits}
\end{minipage}

\end{figure}

The mental health topic shows the most extreme divergence:
80.6\% of VI-associated comments are negative,
versus 42.1\% for HI; HI retains 31.4\% positive
and 26.5\% neutral (Fig.~\ref{fig:mh_sent}).
VI personality profiles show higher neuroticism and lower
conscientiousness than HI (Fig.~\ref{fig:mh_traits}),
consistent with more emotionally reactive and less structured discourse.

\paragraph{Sentiment--trait cross-section.}
\begin{table}[t!]
\centering

\setlength{\tabcolsep}{8pt}
\renewcommand{\arraystretch}{1.25}
\begin{tabular}{l l c c c c c}
\toprule
\textbf{Group} & \textbf{Sentiment} & \textbf{E} & \textbf{N} & \textbf{A} & \textbf{C} & \textbf{O} \\
\midrule
HI & Positive & 0.89 & 0.84 & 0.53 & 0.20 & 0.99 \\
VI & Positive & 0.88 & 0.82 & 0.61 & 0.18 & 0.99 \\
\midrule
HI & Neutral  & 0.84 & 0.93 & 0.43 & 0.20 & 0.99 \\
VI & Neutral  & 0.84 & 0.89 & 0.42 & 0.22 & 0.99 \\
\midrule
HI & Negative & 0.81 & 0.87 & 0.44 & 0.23 & 0.99 \\
VI & Negative & 0.81 & 0.88 & 0.33 & 0.27 & 0.99 \\
\bottomrule
\end{tabular}
\caption{Average Big Five personality trait profiles by sentiment category and influencer type. Traits are abbreviated as follows: E = Extraversion, N = Neuroticism, A = Agreeableness, C = Conscientiousness, O = Openness.}
\label{tab:pos_traits}
\end{table}

Table~\ref{tab:pos_traits} shows that openness is uniformly high
across all conditions, while sentiment-specific differences emerge in
agreeableness, neuroticism, and conscientiousness.
Positive VI comments show the highest agreeableness (affiliative, warm);
negative VI comments show the lowest agreeableness and highest
conscientiousness (critical, deliberate).
This reinforces the two-architecture interpretation:
VI audiences oscillate between warm expressiveness and structured critique,
whereas HI discourse is more uniformly regulated.

\section{Discussion}

\subsection{What virtuality changes in audience discourse}

Across the symbolic, semantic, and topic-level analyses, a consistent picture
emerges: virtual influencer discourse is not simply a shifted version of human
influencer discourse. HI reactions are concentrated around a compact,
stability-centred pattern in which low neuroticism anchors positive sentiment.
VI reactions, by contrast, support multiple discourse regimes, including
appearance-related evaluation, emotionally reactive positivity, and
artificial-identity scrutiny. This suggests that audiences respond to virtual
figures through a broader set of interpretive frames involving not only
admiration, but also questions of authenticity, embodiment, and synthetic
identity.

The appearance finding is especially informative. Although appearance-related
comments occur at nearly the same marginal rate for VI and HI, appearance enters
multiple VI rule clusters and is absent from the HI rule set under the same
thresholds. Thus, the behavioural difference is not merely how often appearance
is discussed, but how appearance becomes structurally embedded with sentiment
and psycholinguistic style.

\subsection{Symbolic--semantic structure of online audience behaviour}

The symbolic--semantic analysis strengthens the behavioural interpretation of
the FCA results. Closed-set mining reveals which affective, topical, and style
signals co-occur systematically; MiniLM embeddings show how these symbolic
configurations form semantic regions. The strong symbolic--semantic alignment
observed for both groups indicates that formal concepts are not arbitrary
attribute bundles: symbolic proximity in the FCA space corresponds to semantic
proximity in embedding space.

At the same time, VI concepts exhibit higher semantic dispersion and a distinct
artificial-identity cluster. This indicates that VI discourse remains organised,
but spans a broader semantic space. For computational social science, this is
important because it shows how symbolic pattern mining and semantic embeddings
can be combined to study not only sentiment polarity, but the organisation of
social meaning in large-scale online discourse.

\subsection{Sensitive topics and artificial identity}

Topic-specific analyses show that VI contexts contain a higher concentration of
negative sentiment in psychologically sensitive domains, especially mental
health, body image, and artificial identity. We interpret these results as
associations between influencer type and discourse organisation, not as causal
evidence that virtuality itself produces negative sentiment. Nevertheless, the
co-occurrence of artificial identity, appearance evaluation, and emotionally
reactive discourse suggests that synthetic embodiment may create distinctive
conditions for audience scrutiny and affective tension.

\subsection{Implications for AI-mediated social interaction}

The results suggest that virtual influencers should not be treated simply as
interchangeable substitutes for human creators. Their synthetic identity appears
to introduce additional interpretive frames through which audiences discuss
authenticity, appearance, and emotional vulnerability. This is particularly
important for topics such as body image and mental health, where VI-linked
discourse shows a higher concentration of negative sentiment. For platforms,
marketers, and creators, these findings suggest that artificial personas should
be deployed cautiously in psychologically sensitive contexts and that audience
responses should be monitored not only through engagement metrics, but also
through the structure of discourse around identity, embodiment, and wellbeing.

\subsection{Ethical considerations}

All analyses are conducted at aggregate level. We do not attempt to identify
commenters, infer protected attributes, or make psychological claims about
individual users. Big Five outputs are treated as textual style indicators rather
than measurements of actual personality. Because user comments may contain
sensitive expressions, especially around mental health and body image, examples
should be paraphrased or reported only in aggregate unless explicit ethical
approval and anonymisation procedures are in place. The goal of the analysis is
to characterise discourse regimes around synthetic social actors, not to evaluate
or profile individual audience members.

\subsection{Limitations}

Several limitations should be considered. First, the data are observational, so
the results should be interpreted as pair-matched structural contrasts rather than
causal effects of virtuality. The VI--HI pairs are matched by niche and subscriber
scale, but unobserved differences in audience demographics, creator style,
moderation practices, posting schedules, and platform recommendation dynamics
may remain. Second, the study is limited to English-language YouTube comments
and three influencer pairs; replication on other platforms and larger cross-niche
samples is needed before generalising to virtual influencers as a whole. Third,
sentiment, topic, and psycholinguistic attributes are inferred automatically and
may contain classification errors. Personality attributes in particular should be
understood as style proxies extracted from text, not as measurements of individual
commenters. Fourth, rule generation relies on an approximate concept-based
mining procedure; association rules are therefore interpreted as statistical
regularities rather than logical implications or causal relationships. Finally,
MiniLM embeddings and cluster labels provide semantic interpretation of formal
concepts, but the underlying evidence remains the closed symbolic structure
extracted from the formal contexts. Future journal extensions should include
manual validation of topic/sentiment labels, independent assessment of cluster
labels, alternative embedding models, and permutation tests for the observed
semantic-dispersion gap.

\section{Conclusion}


We presented a symbolic--semantic computational social science analysis of
audience discourse around virtual and human influencers. Using 69,498 YouTube
comments from three matched VI--HI pairs, we showed that virtuality is
associated with differences not only in sentiment and topic prevalence, but in
the organisation of discourse itself.

Closed-set and rule-based analysis revealed that HI discourse is concentrated
around a compact stability-centred pattern, whereas VI discourse supports
multiple discourse regimes, including an appearance-related pattern absent from
HI despite similar appearance prevalence. Extending this analysis with MiniLM
concept embeddings showed strong symbolic--semantic alignment in both groups,
while VI concepts exhibited higher semantic dispersion and a distinct
artificial-identity cluster.

Together, these findings suggest that virtual influencers reshape online audience
reactions by broadening the semantic space of discourse around identity,
appearance, and emotional engagement. More broadly, the study demonstrates how
symbolic pattern mining and semantic embedding geometry can be combined to
produce interpretable computational social science analyses of large-scale online
behaviour, especially in emerging contexts where synthetic social actors blur the
boundaries between authenticity, performance, and artificial identity.

\bibliographystyle{apalike}
\bibliography{refs}

@book{GanterWille1999,
  author    = {Ganter, Bernhard and Wille, Rudolf},
  title     = {Formal Concept Analysis: Mathematical Foundations},
  publisher = {Springer},
  year      = {1999},
  address   = {Berlin, Heidelberg},
  doi       = {10.1007/978-3-642-59830-2}
}

@article{Stumme2002,
  author    = {Stumme, Gerd and Taouil, Rafik and Bastide, Yves and Pasquier, Nicolas and Lakhal, Lotfi},
  title     = {Computing Iceberg Concept Lattices with {TITANIC}},
  journal   = {Data \& Knowledge Engineering},
  volume    = {42},
  number    = {2},
  pages     = {189--222},
  year      = {2002},
  doi       = {10.1016/S0169-023X(02)00057-5}
}

@inproceedings{Pasquier1999,
  author    = {Pasquier, Nicolas and Bastide, Yves and Taouil, Rafik and Lakhal, Lotfi},
  title     = {Discovering Frequent Closed Itemsets for Association Rules},
  booktitle = {Proceedings of the 7th International Conference on Database Theory ({ICDT~1999})},
  pages     = {398--416},
  year      = {1999},
  publisher = {Springer},
  doi       = {10.1007/3-540-49257-7_25}
}

@article{Poelmans2013,
  author    = {Poelmans, Jonas and Kuznetsov, Sergei O. and Ignatov, Dmitry I. and Dedene, Guido},
  title     = {Formal Concept Analysis in Knowledge Processing: A Survey on Models and Techniques},
  journal   = {Expert Systems with Applications},
  volume    = {40},
  number    = {16},
  pages     = {6601--6623},
  year      = {2013},
  doi       = {10.1016/j.eswa.2013.05.007}
}

@inproceedings{Agrawal1994,
  author    = {Agrawal, Rakesh and Srikant, Ramakrishnan},
  title     = {Fast Algorithms for Mining Association Rules in Large Databases},
  booktitle = {Proceedings of the 20th International Conference on Very Large Data Bases ({VLDB~1994})},
  pages     = {487--499},
  year      = {1994},
  publisher = {Morgan Kaufmann}
}

@article{Arsenyan2022,
  author    = {Arsenyan, Jennifer and Mirowska, Agata},
  title     = {Almost Human? A Comparative Experiment on the Effectiveness of Human and Virtual Influencers},
  journal   = {Psychology \& Marketing},
  volume    = {39},
  number    = {12},
  pages     = {2273--2287},
  year      = {2022},
  doi       = {10.1002/mar.21720}
}

@article{Sands2022,
  author    = {Sands, Sean and Ferraro, Carla and Campbell, Colin and Kietzmann, Jan},
  title     = {Unreal Influence: Leveraging {AI} in Influencer Marketing},
  journal   = {European Journal of Marketing},
  volume    = {56},
  number    = {6},
  pages     = {1721--1747},
  year      = {2022},
  doi       = {10.1108/EJM-12-2019-0949}
}

@article{Lou2022,
  author    = {Lou, Chen and Kiew, Shelly T. J. and Chen, Tao and Lee, Ting Wei and Ong, Jeremy E.-C. and Phua, Joe},
  title     = {Authentically Fake? How Consumers Respond to the Influence of Virtual Influencers},
  journal   = {Journal of Advertising},
  volume    = {52},
  number    = {4},
  pages     = {540--557},
  year      = {2022},
  doi       = {10.1080/00913367.2022.2149641}
}

@inproceedings{Mairesse2007,
  author    = {Mairesse, Fran{\c{c}}ois and Walker, Marilyn A. and Mehl, Matthias R. and Moore, Roger K.},
  title     = {Using Linguistic Cues for the Automatic Recognition of Personality in Conversation and Text},
  booktitle = {Journal of Artificial Intelligence Research},
  volume    = {30},
  pages     = {457--500},
  year      = {2007},
  doi       = {10.1613/jair.2349}
}

@article{yan2024,
  author    = {Ji Yan and Senmao Xia and Amanda Jiang and Zhibin Lin},
  title     = {The Effect of Different Types of Virtual Influencers on Consumers' Emotional Attachment},
  journal   = {Journal of Business Research},
  volume    = {177},
  pages     = {114646},
  year      = {2024},
  publisher = {Elsevier},
  issn      = {0148-2963},
  doi       = {10.1016/j.jbusres.2024.114646},
  url       = {https://www.sciencedirect.com/science/article/pii/S0148296324001504}
}

@article{batista2021,
  author    = {Antonio Batista da Silva Oliveira and Paula Chimenti},
  title     = {"Humanized Robots": A Proposition of Categories to Understand Virtual Influencers},
  journal   = {Australasian Journal of Information Systems},
  volume    = {25},
  year      = {2021},
  publisher = {Australasian Association for Information Systems},
  issn      = {1326-2238},
  doi       = {10.3127/ajis.v25i0.3223},
  url       = {https://doi.org/10.3127/ajis.v25i0.3223}
}

@article{xiecarson2024,
  author    = {Li Xie-Carson and Pierre Benckendorff and Karen Hughes},
  title     = {Keep it \#Unreal: Exploring Instagram Users’ Engagement With Virtual Influencers in Tourism Contexts},
  journal   = {Journal of Hospitality \& Tourism Research},
  volume    = {48},
  number    = {6},
  pages     = {1006--1019},
  year      = {2024},
  publisher = {SAGE Publications},
  issn      = {1096-3480},
  doi       = {10.1177/10963480231180940},
  url       = {https://doi.org/10.1177/10963480231180940}
}

@article{looi2025,
  author    = {Jiemin Looi and Eunjin (Anna) Kim and Zihang E},
  title     = {Sponsorship Disclosure in Virtual Influencer Marketing: Assessing Users’ Sentiment and Engagement Toward Virtual Influencer Endorsements},
  journal   = {Journal of Advertising Research},
  volume    = {0},
  number    = {0},
  pages     = {1--23},
  year      = {2025},
  publisher = {Taylor \& Francis},
  issn      = {0021-8499},
  doi       = {10.1080/00218499.2025.2464300},
  url       = {https://doi.org/10.1080/00218499.2025.2464300}
}
\end{document}